\magnification 1200
\font\pet=cmr8
\rightline{DAMTP/96-56}
\vskip .2 cm
\rightline{SPhT/96-60} 
\vskip 1 cm
\centerline{\bf A RENORMALISATION GROUP STUDY }
\vskip .2 cm
\centerline{\bf OF }
\vskip 0.2 cm
\centerline{\bf THREE DIMENSIONAL TURBULENCE}
\vskip 1 cm
\centerline{by}
\vskip .2 cm
\centerline{Ph. Brax\footnote{$^*$}{\pet
 On leave of absence from SPhT-Saclay
CEA F-91191 Gif sur Yvette, Tel.: (01223) 330853, Fax:(01223) 337918 ,
E-mail:p.brax@amtp.cam.ac.uk }}
\vskip .5 cm
\centerline{DAMTP}
\centerline{University of Cambridge}
\centerline{Silver Street}
\centerline{Cambridge CB39EW UK}
\vskip 3 cm
\leftline{\bf Abstract:}
\vskip .2 cm

We study the three dimensional Navier-Stokes equation with a random
Gaussian
force acting on large wavelengths. Our work has been inspired by
Polyakov's analysis of steady states of two dimensional turbulence. We
investigate 
the time evolution of the probability law of the velocity potential.
 Assuming that this probability law is initially defined by a
statistical field theory in the basin of attraction of a
renormalisation fixed point, we show that 
its time evolution is obtained by averaging over small scale features
of the velocity potential. The probability  law of the velocity potential 
converges to the fixed point in the long
time regime. At the fixed point,
the scaling dimension of the
velocity potential is determined to be  ${-{4\over 3}}$.
We  give conditions for the existence of such a fixed point
of the renormalisation group  describing  the long time behaviour
of the velocity potential. At this fixed point, the energy spectrum of
three dimensional turbulence coincides with a Kolmogorov spectrum.  
\vfill\eject

Turbulence is one of the crucial puzzles of theoretical Physics.
 Despite its practical relevance, there is still no
real thorough understanding of the principles of turbulence. 
A few years ago  Polyakov suggested using conformal field
theories to study turbulence in two dimensions$^{[1]}$.
Polyakov's idea is  that the scale invariant regime of turbulence can be
described by a conformal field theory.
 In three dimensions,  perturbative renormalisation
transformations in momentum space have been used to derive an infrared
fixed point describing the long distance physics of the Navier-Stokes
equation$^{[2,3,4]}$.
Similarly,   probabilistic solutions of the Navier-Stokes
equation with no forcing term 
have been  provided by  self-similar probability laws for
the velocity field$^{[5,6]}$. We shall use properties of the
Navier-Stokes equation under scaling transformations in order to study
the time evolution of solutions. In particular,
our analysis will  combine non-perturbative statements about the
renormalisation group and results about 3d conformal field theories
applied to the Navier-Stokes equation.

The motion of a fluid is described by the macroscopic velocity field.
 We shall suppose that 
the macroscopic velocity field $v_a$ is regularised at small distances
by a cut-off $a$. This cut-off is related to the viscosity by
$$ \nu ={a^2\over \tau}\eqno(1)$$
where $\tau$ is a time characterising the energy decay when a stirring
force is absent. 
 For incompressible
fluids, the velocity field is divergenceless:
$$\nabla .v_a=0.\eqno(2)$$
This implies that the velocity field can be written
$$v_a=\hbox{curl}\  \psi_a\eqno(3)$$
where $\psi_a$ is the velocity potential. We will be interested in the
vorticity
$$\omega_a =\hbox{curl}\ v_a.\eqno(4)$$
The vorticity satisfies the vorticity 
Navier-Stokes equation
$$
{{\partial \omega_a}\over {\partial
t}}+(v_a.\nabla)\omega_a=(\omega_a.\nabla) v_a+\nu\Delta \omega_a +M_a.
\eqno(5)$$
where $M=\hbox{curl} F$ is the vorticity forcing term and $F$ is the
velocity forcing term.
In order to discard boundary effects, we shall only be interested in
the case where the fluid fills up all space. Furthermore, we shall
require the vorticity to decrease at infinity. 
The initial conditions and the external stirring are chosen to be 
regularised random fields where the ultra violet cut-off is given by
the scale $a$.  In particular, the velocity forcing term
will be supposed to be  Gaussian with a white noise temporal
dependence and an action over  large distances.
Taking into account the incompressibility condition,
its 2-point correlation function is 
$$<F_{a,\alpha}(k,\omega)F_{a,\beta}(k',\omega')>=W_0 S_{L}(k)
 (\delta_{\alpha\beta}-{{k^{\alpha}k^{\beta}}\over k^2})    
\delta (k+k')\delta (\omega+\omega')\eqno(6)$$
in Fourier space. The constant $W_0$ determines the amplitude of the
forcing while its $k$-dependence is defined by $S_L(k)$. We suppose
that $S_L$ is a positive smoothing function 
 of width $L^{-1}$, i.e. $S_L(k)=L^3 S(Lk)$
where $S_L$ has  width $L^{-1}$ and is centered around zero, it is
also normalised by $\int d^3 k S(k)=1$. The width is supposed
to be arbitrarily small.
Notice that the spectrum\footnote{$^a$}{\pet
 Practically, 
$ S(k)$ can be taken to
be $ \pi^{-{3\over 2}}\exp -k^2$, in that case $ k^2S_L(k)$ has a
maximum at $ k=L^{-1}$.} of the stirring force $F$ is $k^2S_L(k)$
which is peaked around $k\sim L^{-1}$.
 As $a$ is the only intrinsic scale, the scale $L$ is related to $a$ by
$$L=\kappa a\eqno(7)$$
where $\kappa$ is a large number depending on the forcing term. 
For a given initial configuration and a given realisation of
the stirring force, solutions of (5) will solve the initial value
Cauchy problem.
  Starting from an initial probability law $d{\cal
P}_0(\psi) $, one can solve (5) for a given
realisation of the stirring force. This defines the conditional
probability law
$d{\cal P}(\psi_{a}(x,t)\vert F_a(x,t))$ expressing the
probability law of $\psi_a(x,t)$ for a given realisation of the
stirring force $F_a$.
This probability law is such that the Navier-Stokes equation is valid
in all correlation functions
$$\eqalign{
&{\partial\over \partial t} 
<\omega_a(x_1,t)...\omega_a(x_n,t)>=\cr
&\sum_{i=1}^n<\omega_a(x_1,t)...
(-(v_a.\nabla)\omega_a+(\omega_a.\nabla) v_a+\nu\Delta \omega_a
+M_a)...\omega_a(x_n,t)>\cr}
\eqno(8)$$
for a given realisation of the stirring force $F_a$.
 These equations are the Hopf equations. The conditional probability
law $d{\cal P}(\psi_{a}(x,t)\vert F_a(x,t))$ is random when seen as a
functional of the stirring force.
After averaging over the forcing term
realisations, one gets the probability law of the velocity potential at
time $t$ 
  $$d{\cal P}(\psi_a(x,t))=\int d{\cal
P}(\psi_a(x,t)\vert F_a(x,t))\otimes d\mu (F_a(x,t))\eqno(9)$$
 where
$d\mu (F_a(x,t))$ is the Gaussian probability law of the stirring force
at time $t$. 
 The main object of the
present work is to  characterise the long time behaviour of $d{\cal
P}({\psi_a})$.

We shall suppose that the conditional probability at time $t$ is
defined by a statistical field theory
$$ d{\cal
P}(\psi_a(x,t)\vert F_a(x,t))={1\over Z}d\psi \exp
-S_{t,\tau}(\psi_a,F_a)\eqno(10)$$
where the partition function $Z$ is a normalisation factor, 
$d\psi$ stands for the functional integral symbol and
$S_{t,\tau}(\psi_a,F_a)$ is an effective action at time $t$ depending on the
regularised fields.
By dimensional analysis, the action is a homogeneous function of
$t\over \tau$.
 After integrating over the stirring force, one
obtains the effective action describing the probability law of the
velocity potential at time $t$
$$\eqalign{
d{\cal P}(\psi_a)&={1\over Z}d\psi \exp-S_{t,\tau}(\psi_a)\cr
\exp-S_{t,\tau}(\psi_a)&=\int \exp-S_{t,\tau}
(\psi_a,F_a) d\mu (F_a(x,t))\cr}
\eqno(11)$$
We shall suppose that the velocity potential is a field with scaling
dimension $d_{\psi}$. This dimension is for instance well-defined if
the effective field theory (11) is the  basin of attraction of a fixed
point of the renormalisation group. In that case the scaling dimension
of $\psi$ becomes its conformal dimension.
In this communication, we shall show that the time evolution of solution
of (9) starting from the initial conditions (11) is determined by
renormalisation transformations.
   The
evolution of solutions of (9)
is given by
$$\eqalign{
\tilde\psi_{a}(\lambda x, \lambda^T
t)&=\lambda^{-d_{\psi}}\psi_{\lambda^{-1}a}(x,t)\cr
\exp-S_{\lambda^T t,\tau}(\tilde\psi_a(\lambda x,\lambda^T
t))&=\int_{[{a\over\lambda},a]}
d\psi \exp-S_{t,\lambda^{-T}\tau}(\psi_{\lambda^{-1}a}(x,t)).\cr}
\eqno(12) $$
This is the precise statement that the time evolution of solutions
follows renormalisation trajectories. 
The first equation implies  that the velocity potential lies on  the
renormalisation trajectory of a field of dimension $d_{\psi}$, i.e. 
the solution $\tilde\psi_{a}(\lambda x, \lambda^T t)$ of the
Navier-Stokes equation at time $\lambda^T t$ is simply the renormalised
field $\psi_a(\lambda x,t)$ of $\psi_{\lambda^{-1}a}(x,t)$.   
The second equation entails that
the effective action at time $\lambda^T t$ is obtained after
integrating over the fluctuations in the range
$[\lambda^{-1}a,a]$. This renormalisation transformation is performed
in momentum space integrating over modes in $[a^{-1},\lambda a^{-1}]$. 
The time rescaling exponent $T$ and the scaling dimension of $\psi$
are uniquely determined
$$\eqalign{
T={2\over 3}\cr
d_{\psi}=-{4\over 3}\cr}
\eqno(13)$$
Notice that the scaling dimension of $\omega$ is $d_{\omega}={2\over
3}$ guaranteeing that the vorticity decreases at infinity.

Let us now prove these results. They depend crucially on properties of
the stirring force $F_a$. Notice that the spectrum of the stirring
force $S_L(k)$ has length dimension $-3$. Moreover, one
obtains
$$S_{L}({k\over \lambda})=\lambda^3 S_{\lambda^{-1}
L}(k)\eqno(14)$$
Using (6), one can deduce the following scaling relation
$$<F_a({k\over \lambda},{\omega\over \lambda^T}),
F_a({k'\over \lambda},{\omega'\over
\lambda^T})>=\lambda^{6+T}<F_{\lambda^{-1}a}(k,\omega),
F_{\lambda^{-1}a}(k',\omega')>\eqno(15)$$
As the stirring force is Gaussian, this relation between 2-point
functions yields an equality
$$F_a(\lambda x,\lambda^T t)=\lambda^{-T\over
2}F_{\lambda^{-1}a}(x,t),\eqno(16)$$
 i.e. these two random variables have the same probability
law. 
In other words, one has the equality 
$$\int f(\lambda^{{T\over
2}}F_a(\lambda x,\lambda^T t))d\mu (F_a(\lambda x, \lambda^T t))=
\int f(F_{\lambda^{-1}a}(x,t))d\mu (F_{\lambda^{-1} a}(x,t))\eqno(17)$$
 for any
functional $f$.
This fact will allow us to relate the probability law of the stream
function at two different times when taking the average of (19). 
Let us rewrite the Navier-Stokes equation after rescaling $a\to
\lambda^{-1}a $ and $\tau \to \lambda^{-T} \tau$    
$${ \partial \omega_{\lambda^{-1}a}\over {\partial
t}}+(v_{\lambda^{-1}a}.\nabla)\omega_{\lambda^{-1}a} 
=(\omega_{\lambda^{-1}a}.\nabla)
v_{\lambda^{-1}a}+\nu\lambda^{T-2}\Delta
 \omega_{\lambda^{-1}a} +M_{\lambda^{-1}a}.\eqno(18)$$
At time $t$, the initial condition (10) defined by 
$S_{t,{\tau\over \lambda^T}} 
(\psi_{\lambda^{-1}}(x,t),F_{\lambda^{-1}a}(x,t))$ is a 
solution of  (18) where 
 the range of the integration over  modes for any  correlation function is 
now $[0,\lambda a^{-1}]$.
Using (12) and (16)  one can
substitute $\lambda^{d_{\psi}}\psi_a(\lambda x,\lambda^T t)$ and
$\lambda^{{T\over 2}+1}M_a(\lambda x,\lambda^T t)$ for
$\psi_{\lambda^{-1}a}(x,t)$ and $M_{\lambda^{-1}a}(x,t)$ 
in each correlation function.  
Notice that
 the fields  in  each correlation function have now modes only in the
range $[0,a^{-1}]$. The integration over the modes in the range 
$[a^{-1},\lambda a^{-1}]$ only affects the Boltzmann weight
 $\exp
-S_{t,\lambda^{-T}\tau}
({\psi_{\lambda^{-1}a}(x,t),F_{\lambda^{-1}a}(x,t)})$. Performing
the integration of the Boltzmann weight over these modes gives the
renormalised Boltzmann weight
$$\exp-S_{\lambda^T t,\tau}(\psi_a(\lambda x,\lambda^T
t),\lambda^{{T\over 2}}
F_a(\lambda^T t, \lambda x))=\int_{[{a\over\lambda},a]}
d\psi
\exp-S_{t,{\tau\over\lambda^T}}
(\psi_{\lambda^{-1}a}(x,t),F_{\lambda^{-1}a}(x,t))\eqno(19)$$
The probability law of the velocity potential (12) is obtained
 after averaging over the Gaussian
stirring force and using (17). The resulting equations for the
correlation functions only involves the fields $\psi_a(\lambda
x,\lambda^T t)$ and $M_a(\lambda x,\lambda^T t)$. 
These equations correspond to 
$$
{{\partial \omega_a}\over {\partial
(\lambda^T t)}}+\lambda^{2+d_{\psi}-T}(v_a.\partial_{\lambda x})
\omega_a=\lambda^{2+d_{\psi}-T}(\omega_a.\partial_{\lambda x})
 v_a+\nu\partial^2_{\lambda x} \omega_a +\lambda^{-1-{T\over 2}-d_{\psi}}M_a
\eqno(20)$$
when inserted in correlation functions. The derivatives are taken with
respect to $\lambda^T t$ and $\lambda x$.
 We now require that (20)
coincides with (5) at time $\lambda^T t$ and coordinates
$\lambda x$.      
This is achieved
if
$$\eqalign{
d_{\psi}=-1-{T\over 2}\cr
d_{\psi}=T-2.\cr}\eqno(21)$$  
One can then deduce  (13). This proves that (12) and (13) specify the time
evolution of the probability law of the velocity potential.

Starting from a field theory in the basin of attraction of a fixed
point of the renormalisation group, this probability follows
renormalisation trajectories. It therefore converges to the
probability law specified by the fixed point. 
The scaling dimension $-{4\over 3}$ of the velocity potential
becomes  its conformal dimension at the fixed point ( fixed
points of the renormalisation group are conformal theories$^{[7]}$). 
We shall now derive further  conditions for the existence of such a fixed
point.
As the long time regime is linked to the small $a\over \lambda$ region
by (12), the fixed points satisfy (18) in the limit when $\lambda$
goes to infinity.
Using (13), one can see that the influence of the viscosity becomes
negligible in the long time regime as $T-2=-{4\over 3}$. Similarly,
the stirring force can be easily dealt with. Indeed, notice that
$$S_{L\over \lambda}(k)= ({L\over \lambda})^3S ({kL\over
\lambda})\eqno(22)$$ 
converges to zero as $\lambda \to \infty$. This implies that the
2-point correlation of $F_{\lambda^{-1}a}$ goes to zero and therefore
the Gaussian stirring force vanishes in the long time
regime. Assuming that the fixed points represent steady states, the
time derivatives of each correlation function vanish as well. We
therefore find that fixed points are characterised by the vanishing of the
non-linear terms in the limit $\lambda\to \infty$.   
 The non-linear terms can be readily
evaluated when the cut-off is rescaled to zero using properties of
short distance expansions.   
 Products of fields become singular when the cut-off is
removed. For fields called quasi-primary fields$^{[8]}$,
the result of the product  can  be expanded in 
 a power series
in   ${a\over \lambda}$. A vanishing limit is obtained if the leading
term has a positive exponent.
We shall suppose that the velocity potential becomes a quasi-primary
field at the fixed point. In that case, 
the nonlinear terms read
$$\epsilon_{\alpha\beta\gamma}\partial_{\gamma}
((v_{\lambda^{-1}a}.\nabla)v_{\lambda^{-1}a})_{\beta}\sim
({a\over \lambda})^{d_2-4-2d_{\psi}}
\psi_{2,\alpha,\lambda^{-1}a}+...\eqno(23)$$
where $d_2$ is the conformal dimension of the leading pseuso-vector
$\psi_{2,\alpha}$ in the expansion of the non-linear terms.
 We can therefore conclude that the fixed points are
characterised by the inequality
$$d_2>4+2d_{\psi}\eqno(24)$$
This generalises a similar inequality obtained by Polyakov in two
dimensions. We can now state our main result.
Conformal field theories satisfying (24) such that the velocity potential
is a  quasi-primary field of dimension $-{4\over3}$ describe the long
time regime of 3d turbulence.

As a consequence, 
notice that $\psi$ cannot be identified with $\psi_2$ as (24) is not
satisfied in that case. This implies that 
$$<\psi_{\mu}\psi_{\nu}\psi_{\rho}>=0,\eqno(25)$$
the three point function of the velocity potential vanishes identically
at the fixed point.
It is extremely difficult to construct explicitly three dimensional
conformal field theories describing an infrared fixed point of the
renormalisation group. We shall suppose that such
theories exist and deduce consequences on the energy spectrum of the
solutions of the Navier-Stokes equation. 
In particular, we can calculate the 2-point
function of the velocity potential and then the energy spectrum in the
long time regime.
The energy spectrum is easily related to the Fourier transform of the
two point function
$$E_a(k)=4\pi k^2\int d^3x <v_a(x) v_a(0)>\exp -2\pi ik.x.\eqno(26)$$  
In the long time regime, the energy spectrum behaves as 
$$E_a(k)\sim k^{-{5\over 3}}.\eqno(27)$$
This  spectrum is valid for $k\ll {1\over a}$.
The exponent is given by $2d_{\psi}+1=-{5\over 3}$
\footnote{$^b$}{\pet Our results are compatible with the perturbative
calculations of Refs. [2,3,4]. In this case, the 
 stirring force
is specified by
$<F_{a,\alpha}(k,\omega)F_{a,\beta}(k',\omega')>={W_0\over (k^2
+m^2)^{y\over 2}} 
\chi_{a^{-1}}(k)
(\delta_{\alpha\beta}-{{k^{\alpha}k^{\beta}}\over k^2})    
\delta (k+k')\delta (\omega+\omega')$
where $\chi_{a^{-1}}(k)$ is equal to one for $k\le a^{-1}$ and zero
otherwise.
The mass term $m=L^{-1}$ is small but guarantees that there are no infrared
divergences when $k\to 0$. The same analysis as in the text shows that
$d_{\psi}=-{{1+y}\over 3}$and  the spectrum behaves like
$k^{{1-2y}\over 3}$ in the range $k\ll a^{-1}$ for $0<y\le 5$.}.  
The spectrum coincides with a Kolmogorov spectrum$^{[9]}$

The solutions of the Navier-Stokes equation (12)
 have
some salient features. Let us notice that 
 the evolution of the probability law of the velocity potential (12)
makes explicit the breaking of time reversal invariance already
present in  the
Navier-Stokes equation. Indeed, the
probability law at time $\lambda^T t$ is obtained from the probability
law at time $t$ by averaging over the small scale features of the
velocity potential. These small scale features are therefore lost in the
process of evolution. In particular, knowing the probability law at
time $\lambda^T t$ does not enable one  to deduce the probability law at
time $t$. Another consequence is 
 that  the renormalisation trajectories converge to a
non-unitary conformal field theory as $d_{\psi}<0$. This result has
also been obtained in 2d$^{[1]}$. It reinforces Polyakov's idea that
turbulence is a flux state and not an equilibrium state described by
an unitary theory. Moreover, at the fixed point the energy spectrum is 
determined to be a Kolmogorov spectrum. This result highly depends
on the fact that the stirring force is Gaussian and  acts on large
wavelengths.  Another feature of our analysis is that we
have derived a spectrum compatible with a Kolmogorov spectrum without
any  cascade hypothesis$^{[9]}$.
Finally, let us note that our results can be applied in two
dimensions, in particular (24) is replaced by Polyakov's
condition$^{[1]}$\footnote{$^c$}{
\pet The stream function $\psi$ satisfies $\psi_{\lambda^{-1}a}(x)
\psi_{\lambda^{-1}a}(x)\sim ({a\over
\lambda})^{d_2-2d_{\psi}}\psi_{2,\lambda^{-1}a}(x) $ with $d_2>
2d_{\psi}$.}
and the energy spectrum is still given by a Kolmogorov
spectrum.

We have been able to obtain properties of solutions of the 3d
Navier-Stokes equation in the long
time  regime.  Assuming that the probability law of the stream
function is a statistical field theory, its time evolution  corresponds to
renormalisation trajectories. Starting from initial conditions in the
basin of attraction of a fixed point under renormalisation
transformations, the solutions converge asymptotically to the fixed
point. The possible fixed points are all non-unitary. 
  In the case where
the random stirring has a spectrum concentrated on large  wavelengths,
we have shown that a Kolmogorov spectrum is obtained in the long
time regime. The existence of such a fixed point of the
renormalisation group governing the long time regime of turbulence
seems physically clear in view of the numerous experimental as well as
computational evidence in favour of a Kolmogorov spectrum. It is a
challenging problem to construct such a field theory.

\noindent acknowledgements: 

I am grateful to W. Eholzer, M. Gaberdiel, H. K. Moffatt and
R. Peschanski for useful discussions and suggestions. 
 
\vfill\eject
\vskip 1cm
\leftline{\bf References}
\vskip 1 cm
\vskip .2 cm
\leftline{[1] A. M. Polyakov Nucl. Phys. {\bf B396} (1993) 397}
\vskip .2 cm
\leftline{[2] D. Forster, D. R. Nelson and M. J. Stephen Phys. Rev.
Lett {\bf 36} (1976) 867}
\vskip .2 cm
\leftline{[3] D. Forster and D. M. Nelson Phys. Rev. {\bf A16} (1977)
732}
\vskip .2 cm
\leftline{[4] C. Dedominicis and P. C. Martin Phys. Rev. {\bf A19} (1979) 419}
\vskip .2 cm
\leftline{[5] C. Foias and R. Temam Comm. Math. Phys. {\bf 90} (1983)
181}
\vskip .2 cm
\leftline{[6] C. Foias and R. Temam India. Univ. Math. J. {\bf 29}
(1980) 913}
\vskip .2 cm
\leftline{[7] A.A. Belavin, A. M. Polyakov and A. B. Zamolodchikov
Nucl. Phys {\bf B 241} (1984) 333}
\vskip .2 cm
\leftline{[8] S. Ferrara, A. F. Grillo and R. Gatto Annals of
Phys. {\bf 76} (1973) 161}
\vskip .2 cm
\leftline{[9] A. N. Kolmogorov C. R. Acad. Sci. URSS {\bf 30} (1941)
301}
\end